\documentclass[11pt]{article}
\usepackage[final]{acl}
\usepackage{times}
\usepackage{latexsym}
\usepackage[T1]{fontenc}
\usepackage[utf8]{inputenc}
\usepackage{microtype}
\usepackage{graphicx}
\usepackage{booktabs}
\usepackage{array}
\usepackage{amsmath,amssymb}
\allowdisplaybreaks 
\usepackage{url}
\makeatletter\g@addto@macro\UrlBreaks{\do\-}\makeatother 

\title{URCHIN: A Horizontal Spiking Language Model for Data-Constrained Pretraining}
\author{Po-Han Chiang \\[2pt]
  \normalfont\small
  \begin{minipage}{0.92\textwidth}\centering
  National Yang Ming Chiao Tung University \\
  \texttt{phc@nycu.edu.tw}
  \end{minipage}}

\begin{document}
\maketitle

\begin{abstract}
The BabyLM challenge measures how much language a model can learn from developmentally-plausible, child-scale data rather than internet-scale corpora, yet prior language models forgo the biological constraints of the neural circuitry that acquires human language: spiking neurons separated into excitatory and inhibitory populations wired by a recurrent lateral connectome. This paper presents URCHIN (Unified Recurrent Connectome with Horizontal Integrate-and-fire Neurons), which applies the Parallelized Hierarchical Connectome Spiking State-space Model (PHCSSM) to language modeling: leaky integrate-and-fire neurons coupled by a Dale's-law lateral connectome resolve each token through a multi-transmission loop that recirculates activity to a fixed point. The instantiation is deliberately minimal: a single horizontal layer of 128 neurons, no attention, and 4.23M parameters. Two implementations share one set of weights and produce identical benchmark scores, so URCHIN is trained once and deployed either way with no conversion step: a parallel state-space model (SSM) scan that is GPU-efficient for training, or an event-driven recurrent spiking neural network (RSNN) with constant-cost inference for CPU or neuromorphic edge deployment. Across all three BabyLM tracks (Strict-100M, Strict-Small, and Multilingual), URCHIN offers a biologically plausible, efficient, and directly deployable reference point.
\end{abstract}

\section{Introduction}

Scaling to internet-scale corpora has driven most recent progress in language modeling, but it has also moved these models far from the conditions under which humans acquire language. A child reaches broad linguistic competence from on the order of ten to a hundred million words, and the BabyLM challenge takes this as its benchmark, capping training data at a developmentally-plausible budget instead of allowing internet-scale text. The gap this exposes is not only about the amount of data. The dominant recipe, which the official BabyLM baselines \citep{radford2019,charpentier2024} and most entries \citep{charpentier2025} follow, is a feedforward transformer \citep{vaswani2017}: a stack of roughly a dozen attention layers that processes each token in a single forward pass. In humans, by contrast, language is learned and run by the brain, whose circuits differ from these networks in ways usually treated as biological detail rather than as design guidance: neurons communicate through discrete spikes, each obeys Dale's law and is purely excitatory or purely inhibitory, connectivity within a region is recurrent and lateral rather than feedforward, and the system as a whole operates at a tiny energy budget. Spiking neural networks that adopt these constraints are a rich and active research area, yet they have largely stayed out of language modeling for a practical reason: spikes are event-driven and inherently sequential, so such networks are hard to parallelize and slow to train on the token volumes that language modeling requires. This paper asks whether that barrier can be crossed cheaply: can a minimal, biologically-constrained spiking network learn language under the BabyLM budget, and what does it cost to train and to deploy?

The model introduced here is URCHIN (Unified Recurrent Connectome with Horizontal Integrate-and-fire Neurons).\footnote{The name echoes the spiny sea urchin, a fitting image for a spiking network of only a few million parameters.} It adapts a recent architecture, the Parallelized Hierarchical Connectome Spiking State-space Model (PHCSSM) of \citet{chiang2026}, which folds a set of biological priors into a trainable spiking network and can be evaluated either as a state-space model (SSM) with parallel scans or as an event-driven recurrent spiking neural network (RSNN), giving the same result both ways. The parallel form is what removes the training bottleneck, so a biologically-constrained spiking network can be trained on language-model-scale data, while the serial form preserves the event-driven, energy-frugal behavior for deployment. URCHIN carries PHCSSM over to autoregressive language modeling: leaky integrate-and-fire (LIF) neurons coupled only by lateral connections, where a token is resolved not in one forward step but by a multi-transmission loop that recirculates activity through those connections, reusing one shared synapse and neuron nonlinearity at every step until it settles to a fixed point. The name records the two properties that define the method: a unified recurrent connectome, a single weight set shared between training and deployment, and horizontal, lateral-only connectivity. Nothing in this design fixes the number of layers, and the entry to the challenge is deliberately the smallest useful instance, a single horizontal layer of 128 neurons. The minimalism is deliberate: earlier spiking language models (LMs) such as SpikeGPT, SpikeLM, SpikeLLM, and SpikingBERT are surrogate-gradient spiking transformers or RWKV backbones, trained at scale, without biological connectivity constraints, and outside the BabyLM setting; URCHIN instead keeps the biological constraints and asks how far a single tiny recurrent layer can go.

URCHIN is positioned as a contribution to efficiency and biological plausibility. The contribution is threefold.

First, this paper presents a minimal spiking language model with biological constraints in the BabyLM setting. URCHIN is the first spiking network with a Dale's-law excitatory/inhibitory (E/I) connectome and an event-driven deployment mode that is trained and evaluated under the BabyLM data constraints across all three 2026 tracks, and it does so here in a minimal single-layer form in which the same neurons carry the input and the output and are joined only by lateral connections.

Second, this paper shows train and deploy score equivalence for a spiking language model. Parallel URCHIN (the parallel-scan SSM training backbone) and Serial URCHIN (the event-driven RSNN deployment backbone) share one set of weights and give the same benchmark scores across the full evaluation suite, agreeing to the floating-point floor of the two integration orders. This property was established for PHCSSM in the time-series setting, and is confirmed here to carry over to language modeling. It is not the same as ANN-to-SNN (artificial neural network to spiking neural network) conversion equivalence, because URCHIN is natively spiking and needs no conversion step.

Third, URCHIN is shown to be efficient in both parameters and compute. The spiking core is a 34K-parameter single-layer recurrent circuit, which is 0.8 percent of the 4.23M total; the rest is the token embedding and the output projection. Pretraining costs about 15 to 50 times less compute than the GPT-2 and GPT-BERT baselines.

\section{Related Work}

\paragraph{Spiking language models.} SpikeGPT \citep{zhu2023} introduces a generative RWKV-style spiking language model with binary activations. SpikeLM \citep{xing2024a} uses bidirectional ternary firing. SpikeLLM \citep{xing2024b} scales spiking computation toward large language models. SpikingBERT \citep{bal2024} targets encoder-style text classification, and SpikingBrain 2.0 \citep{pan2026} studies brain-inspired foundation models. These systems show that spiking computation is compatible with language modeling. They use spiking transformers or RWKV backbones trained with surrogate gradients (a smooth approximation that lets the non-differentiable spike be backpropagated), without lateral recurrent connectivity between neurons or biological connectivity constraints, they are trained without a developmental data budget, and they are not BabyLM entries.

\paragraph{Spiking state-space models.} SpikingSSMs \citep{shen2025} apply sparse spiking state-space models to long sequences, with a diagonal state-space recurrence in which neurons carry no lateral connections to one another. URCHIN differs on several points at once: its neurons are joined by a lateral recurrent connectome, it is biologically constrained through Dale's law on that connectome, and it is trained under the BabyLM data budget.

\paragraph{ANN-to-SNN conversion.} A separate line of work trains a conventional rate-coded ANN and transfers its weights to a spiking network that emulates the ANN's continuous activations through firing rates over a deployment-time window \citep{cao2015,rueckauer2017,sengupta2019,bu2022}. This inherits the source ANN's parallel-training scalability but at the cost of a conversion-stage accuracy gap and an emulation window of tens to hundreds of timesteps. URCHIN is instead natively spiking and trained directly in the spiking regime; its parallel and serial modes are two views of a single weight set, so its train-and-deploy equivalence needs no conversion step and incurs no conversion gap.

\paragraph{BabyLM.} The challenge \citep{choshen2026} and its GPT-2 \citep{radford2019} and GPT-BERT \citep{charpentier2024} baselines define the data budget and the evaluation suite used here. Most submissions are transformers, but a line of work explores non-transformer and subquadratic architectures for sample-efficient pretraining, such as the recurrent BabyHGRN \citep{haller2024}; URCHIN extends this line to a biologically-constrained spiking model.

\paragraph{PHCSSM.} The model introduced in Chiang (2026) is the first spiking SSM to combine five biological constraints: lateral (horizontal) connectivity, Dale's law, adaptive leaky integrate-and-fire dynamics (ALIF), short-term plasticity (STP), and spike-timing-dependent plasticity (STDP). It is also the first SSM whose single set of weights runs in both a parallel-scan mode for training and a serial event-driven RSNN mode for deployment, with no ANN-to-SNN conversion. It was evaluated only on the UEA MTSCA (the multivariate time-series classification archive), over signal domains such as EEG, ECG, spectral, and worm-motion data, and had not been used for language modeling. URCHIN is its first application to language, following the common BabyLM practice of adapting an established architecture to the challenge; Section~\ref{sec:model} describes what carries over from PHCSSM and what changes.

To the best of the author's knowledge, no biologically-constrained spiking language model of this form, in which the neurons communicate by discrete spikes, are each exclusively excitatory or inhibitory, and are wired by a recurrent lateral connectome, has been trained and evaluated under the BabyLM data budget before URCHIN.

\section{The URCHIN System}
\label{sec:model}

\begin{figure}[t]
\centering
\includegraphics[width=77mm]{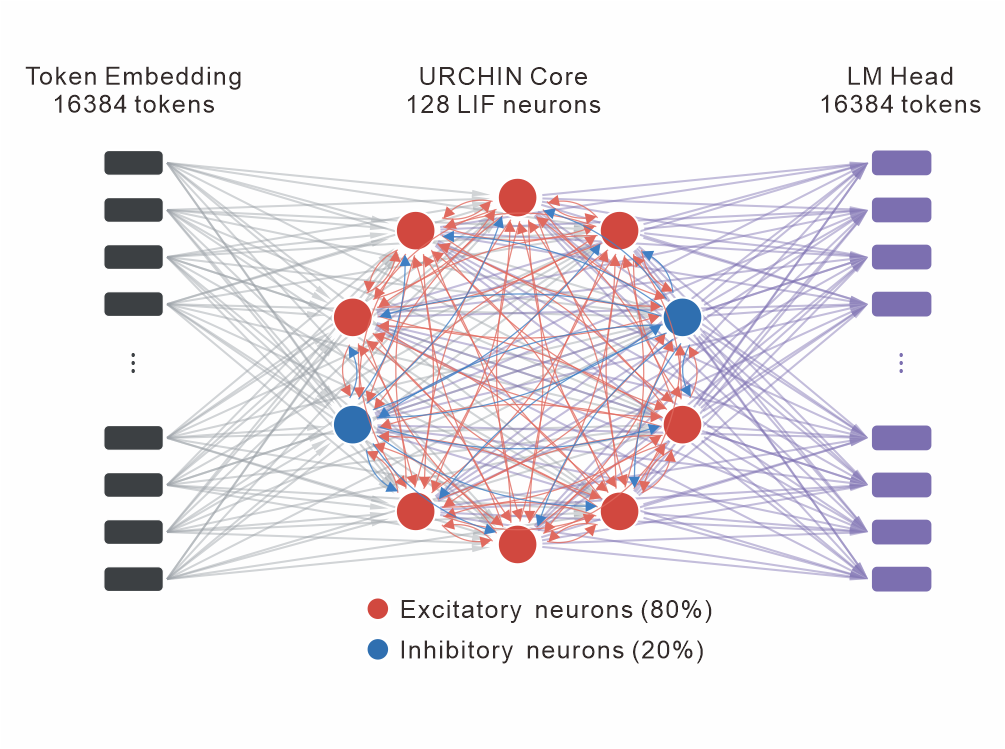}
\caption{URCHIN architecture. The embedding of the current token is injected as input current into a single layer of excitatory (red) and inhibitory (blue) LIF neurons joined only by lateral connections; a linear LM head reads the membrane voltage of the same neurons out to the 16,384-entry vocabulary. Each token is processed by a multi-transmission loop, a fixed number of lateral transmission steps that settles to a fixed point. Parallel URCHIN evaluates this with parallel scans for training; Serial URCHIN runs the identical dynamics event-driven for deployment.}
\label{fig:arch}
\end{figure}

URCHIN is the language-modeling instance of PHCSSM, reduced to a single horizontal layer. This section describes the components that matter for the submission; the full neuron and synapse equations, with their decay and reset terms and the parallel-scan derivation, are given by PHCSSM \citep{chiang2026}. The submitted model is a single configuration whose architecture is unchanged across all three tracks.

\paragraph{Relation to PHCSSM.} URCHIN uses the same architecture as PHCSSM. The LIF neurons, the Dale's-law E/I synapse, the lateral connectome, and the within-token multi-transmission recurrence (with its parallel-scan and event-driven implementations) are shared by both models. What changes is the input and output: in place of the signal encoder and classification head that PHCSSM used for time-series classification, URCHIN feeds the token embedding as input current and reads the membrane voltage through a linear LM head, trained with a causal next-token objective. Within the core, the submitted model uses only plain LIF neurons; the ALIF, STP, and STDP functions are not used here.

\paragraph{A single horizontal layer.} The model is one recurrent region of 128 LIF neurons \citep{bellec2018}, with no depth and no region hierarchy (Figure~\ref{fig:arch}). A LIF neuron sums its input into a membrane voltage that leaks toward a resting value over time; when the voltage reaches a threshold the neuron emits a spike and the voltage resets. The token embedding is injected as input current into these neurons, and the LM head reads the membrane voltage of the same neurons, so a single population carries both the input and the output. The neurons are connected only by horizontal (lateral) connections; there are no feedforward layers between an input population and an output population, because there is only one population.

\paragraph{Dale's law and E/I structure.} The 128 neurons are split into excitatory and inhibitory populations, and synaptic weights obey Dale's law \citep{cornford2021,balwani2025}: a sign constraint holds each neuron's outgoing projections purely excitatory or purely inhibitory throughout training and inference. The lateral connectivity is full over the four E and I blocks, \texttt{[E->E, I->E, E->I, I->I] = [1,1,1,1]}. All three tracks use the same homogeneous population with an excitatory-to-inhibitory ratio of 80 to 20 and a single membrane time constant.

\paragraph{Multi-transmission loop and readout.} For each input token the layer runs a fixed number of lateral transmission steps over its horizontal connections, a within-token loop that repeats until the state stops changing. URCHIN uses the plain-LIF forward pass of PHCSSM \citep{chiang2026}. At step $k$ and token position $t$, the token embedding $e_t$ enters as input drive with gain $\alpha_{\mathrm{drive}}$, and each neuron sends its previous spikes back through the Dale-masked lateral weight $W_{\mathrm{struct}} = W_{\mathrm{syn}} \odot M_{\mathrm{topo}}$ ($W_{\mathrm{syn}}$ sign-clamped so each neuron's outgoing weights keep one sign, $M_{\mathrm{topo}}$ the fixed lateral-topology mask, $d$ the synaptic delay). Each neuron carries an excitatory-membrane state $V_{\mathrm{exc}}$ and a refractory-reset state $V_{\mathrm{res}}$ with learnable leak factors $\alpha_{\mathrm{exc}}, \alpha_{\mathrm{ref}} \in (0,1)$, applies the softplus nonlinearity $\zeta(x) = \log(1+e^{x})$, and fires through the Heaviside $H$ at threshold $V_{\mathrm{th}}$, with $s^{\mathrm{pre}}_t = H(V_{\mathrm{exc},t} - V_{\mathrm{th}})$ the preliminary crossing. For $k = 1, \dots, K$, with the neuron states reset at the start of each step and $s^{(0)}=0$,
\begin{gather}
I^{(k)}_t = W_{\mathrm{struct}}\, s^{(k-1)}_{t-d} + \alpha_{\mathrm{drive}}\, e_t, \\
V_{\mathrm{exc},t} = \alpha_{\mathrm{exc}} V_{\mathrm{exc},t-1} + \zeta(I^{(k)}_t), \\
V_{\mathrm{res},t} = \alpha_{\mathrm{ref}} V_{\mathrm{res},t-1} + \zeta(w_{\mathrm{reset}}\, s^{\mathrm{pre}}_t), \\
V_{\mathrm{mem},t}^{(k)} = V_{\mathrm{exc},t} - V_{\mathrm{res},t}, \\
s^{(k)}_t = H(V_{\mathrm{mem},t}^{(k)} - V_{\mathrm{th}}).
\end{gather}
The membrane voltage $V_{\mathrm{mem}} = V_{\mathrm{exc}} - V_{\mathrm{res}}$ is the excitatory-membrane state minus the refractory reset, and the spikes $s^{(k)}$ of one step become the lateral input to the next. The loop settles to a within-token fixed point $s^{(k)} = s^{(k-1)}$; training uses 13 transmission steps (a target of 12 plus one buffer step) and evaluation uses 24, since additional steps at inference move the state closer to that fixed point. State is reset for each token, so this loop is a within-token recurrence, separate from the sequence axis, which is evaluated as a log-domain parallel prefix scan over the token positions. A linear LM head then reads the final membrane voltage of the same neurons out to the 16,384-entry vocabulary (the byte-pair-encoding (BPE) tokenizer provided by the BabyLM challenge),
\begin{equation}
\hat{y}_t = W_{\mathrm{dec}}\, V_{\mathrm{mem},t}^{(K)} + b_{\mathrm{dec}},
\end{equation}
with the head weights $W_{\mathrm{dec}}, b_{\mathrm{dec}}$ untied from the token embedding. Training uses a causal next-token objective and backpropagation only.

\paragraph{Parallel URCHIN (SSM) and Serial URCHIN (RSNN).} URCHIN has two implementations that share a single set of weights and are two mathematically equivalent views of the same dynamics. Parallel URCHIN is the SSM form: it casts the multi-transmission loop and the sequence recurrence as linear state-space recurrences and evaluates them in JAX \citep{bradbury2018} and PyTorch with parallel prefix scans, so it trains with backpropagation at transformer-like throughput; it is used for all training. Serial URCHIN is the RSNN form: it runs the identical dynamics event-driven, one timestep at a time, and is the deployment target for neuromorphic and edge hardware. The two forms are the same model up to floating-point summation order, and Section~\ref{sec:results} reports their agreement.

\paragraph{Parameter budget.} The submitted model has about 4.23M stored parameters, of which about 4.21M are trainable; the difference is the fixed Dale's-law connectivity mask, which is stored but not learned (Table~\ref{tab:params}). The token embedding and the untied LM head account for 99.2 percent of the total (two blocks of 16384 by 128). The spiking core is about 34K parameters (about 17K trainable; the remainder is the fixed Dale mask, Table~\ref{tab:params}), so URCHIN performs language modeling with a single 34K-parameter recurrent core.

\begin{table}[t]
\centering
\small
\begin{tabular}{@{}lrr@{}}
\toprule
Component & Trainable & Total (stored) \\
\midrule
Token embedding & 2.10M & 2.10M \\
LM head & 2.10M & 2.10M \\
Spiking core & 17K & 17K \\
Fixed Dale + input masks & 0 & 16.6K \\
\midrule
\textbf{Total} & \textbf{4.21M} & \textbf{4.23M} \\
\bottomrule
\end{tabular}
\caption{Parameter budget of the submitted model. Trainable parameters against total stored tensors. The fixed Dale's-law connectivity mask is stored in the checkpoint but is not trained.}
\label{tab:params}
\end{table}

\section{Experimental Setup}

\paragraph{Tracks and data.} The same architecture is submitted to all three tracks of the BabyLM challenge: Strict-Small (10M words), Strict-100M (100M words), and Multilingual (100M words in English, Dutch, and Chinese) drawn from BabyBabelLM \citep{jumelet2025}. The strict tracks use the official BabyLM corpus, drawn from sources including CHILDES \citep{macwhinney2000}, Project Gutenberg \citep{gerlach2018}, OpenSubtitles \citep{lison2016}, and the Switchboard Dialog Act Corpus \citep{stolcke2000}. Each track uses the tokenizer that the BabyLM challenge provides for it: a BPE tokenizer with a 16,384 vocabulary in all three cases, but a different tokenizer per track. The vocabulary size is therefore shared, while the token inventories differ; the Multilingual tokenizer covers English, Dutch, and Chinese and shares only about a quarter of its tokens with the English strict-track tokenizers. The full pretraining and fine-tuning hyperparameters for all three tracks are given in Appendix~\ref{app:hyper} (Table~\ref{tab:hyper}); the architecture is identical across tracks, and only the corpus, its tokenizer, and the schedule differ. Training uses a causal objective and backpropagation only, and stays within the BabyLM training budget (a fixed cap on epochs and tokens set by the challenge). The three submitted checkpoints (\texttt{URCHIN\_BabyLM2026\_StrictSmall}, \texttt{URCHIN\_BabyLM2026\_Strict}, and \texttt{URCHIN\_BabyLM2026\_Multilingual}), each packaging both the parallel and serial implementations under one set of weights, are publicly available on the Hugging Face Hub (\url{https://huggingface.co/phclab}).

\paragraph{Evaluation.} Evaluation uses the official evaluation pipeline released by the BabyLM challenge (\texttt{babylm-eval}, commit 3d57ddc). It covers BLiMP \citep{warstadt2020}, BLiMP-Supplement, EWoK \citep{ivanova2024}, Entity-Tracking \citep{kim2023}, COMPS \citep{misra2023}, GlobalPIQA, reading-behavior fits, and the GLUE \citep{wang2019} fine-tuning suite, together with the multilingual suite for the ML track.

\paragraph{Fine-tuning.} Fine-tuning uses the challenge pipeline with its optional \texttt{-{}-lr} flag set to 3e-4 for the strict-track (Super)GLUE tasks and 1e-3 for the multilingual suite (Table~\ref{tab:hyper}), staying within the epoch budget under best-per-epoch early stopping.

\section{Results}
\label{sec:results}

The same homogeneous single-layer model is submitted to all three tracks; they share one architecture and differ only in the training corpus and the schedule.

\begin{table}[t]
\centering
\small
\begin{tabular}{@{}lrrr@{}}
\toprule
Metric & URCHIN & \shortstack{Baseline-\\GPT2} & \shortstack{Baseline-\\Interaction} \\
\midrule
Overall Average & 37.01 & 37.38 & 38.71 \\
BLiMP & 61.29 & 65.23 & 63.07 \\
BLiMP-Supplement & 56.05 & 57.25 & 58.13 \\
EWoK & 50.29 & 50.63 & 51.45 \\
Entity-Tracking (deb.) & 18.16 & 19.10 & 19.62 \\
COMPS & 50.70 & 51.81 & 51.29 \\
GlobalPIQA & 29.70 & 35.09 & 36.14 \\
(Super)GLUE & 59.47 & 63.80 & 63.62 \\
Reading (combined) & \textbf{7.43} & 5.63 & 5.10 \\
\quad Eye-Tracking & 10.00 & 9.63 & 8.50 \\
\quad Self-paced Reading & 4.85 & 1.64 & 1.69 \\
AoA (official) & \textbf{0.0} & $-12.15$ & 0.0 \\
NLP Average & 46.52 & 48.99 & 49.04 \\
Human-like Average & \textbf{3.71} & $-3.26$ & 2.55 \\
\bottomrule
\end{tabular}
\caption{Strict-Small official results: URCHIN against the two official BabyLM baselines, the GPT-2 baseline (\url{BabyLM-2026-Baseline-GPT2-Strict-Small}) and the interaction baseline (\url{BabyLM-2026-Baseline-Strict-Small-Interaction}). Bold marks the metrics on which URCHIN matches or exceeds both baselines. AoA is decoupled to zero in URCHIN by design.}
\label{tab:strictsmall}
\end{table}

\begin{table}[t]
\centering
\small
\begin{tabular}{@{}lrrr@{}}
\toprule
Metric & URCHIN & \shortstack{Baseline-\\GPT2} & \shortstack{Baseline-\\Interaction} \\
\midrule
Overall Average & 37.54 & 40.73 & 39.89 \\
BLiMP & 63.08 & 74.73 & 72.89 \\
BLiMP-Supplement & 52.57 & 65.00 & 65.09 \\
EWoK & 49.97 & 54.37 & 54.23 \\
Entity-Tracking (deb.) & \textbf{18.39} & 16.91 & 15.89 \\
COMPS & 52.73 & 55.85 & 55.37 \\
GlobalPIQA & 32.74 & 36.62 & 36.18 \\
(Super)GLUE & 61.19 & 67.75 & 67.28 \\
Reading (combined) & \textbf{7.18} & 6.93 & 3.43 \\
\quad Eye-Tracking & 10.28 & 10.54 & 4.74 \\
\quad Self-paced Reading & 4.08 & 3.32 & 2.13 \\
AoA (official) & \textbf{0.0} & $-11.58$ & $-11.36$ \\
NLP Average & 47.24 & 53.03 & 52.42 \\
Human-like Average & \textbf{3.59} & $-2.32$ & $-3.96$ \\
\bottomrule
\end{tabular}
\caption{Strict-100M official results: URCHIN against the two official BabyLM baselines, the GPT-2 baseline (\url{BabyLM-2026-Baseline-GPT2-Strict}) and the interaction baseline (\url{BabyLM-2026-Baseline-Strict-Interaction}), formatted as in Table~\ref{tab:strictsmall}. Bold marks the metrics on which URCHIN matches or exceeds both baselines.}
\label{tab:strict100}
\end{table}

\begin{table}[t]
\centering
\small
\begin{tabular}{@{}lrrrr@{}}
\toprule
 & En & Nl & Zh & Overall \\
\midrule
\multicolumn{5}{@{}l}{\textbf{URCHIN}} \\
\quad Zero-shot avg. & 48.60 & 48.72 & 46.70 & 48.01 \\
\quad Fine-tune avg. & 39.28 & 39.55 & \textbf{41.47} & 40.10 \\
\quad Language avg. & 43.27 & 43.83 & \textbf{43.57} & 43.56 \\
\midrule
\multicolumn{5}{@{}l}{\textbf{Baseline-GPT2 (en/nl/zh)}} \\
\quad Zero-shot avg. & 53.63 & 55.05 & 47.25 & 51.98 \\
\quad Fine-tune avg. & 43.09 & 39.82 & 40.69 & 41.20 \\
\quad Language avg. & 47.60 & 46.92 & 43.31 & 45.94 \\
\midrule
\multicolumn{5}{@{}l}{\textbf{Baseline-GPT2 (multilingual)}} \\
\quad Zero-shot avg. & 35.42 & 49.04 & 41.71 & 42.06 \\
\quad Fine-tune avg. & 31.51 & 29.76 & 31.68 & 30.98 \\
\quad Language avg. & 33.18 & 38.76 & 35.69 & 35.88 \\
\bottomrule
\end{tabular}
\caption{Multilingual official results: per-language zero-shot, fine-tune, and language averages for URCHIN and the two official GPT-2 baselines, \url{BabyLM-2026-Baseline-GPT2-en_nld_zho_equal} (en/nl/zh) and \url{babylm-multilingual-gpt2-baseline} (multilingual). All figures are from the official leaderboard. The Overall column is the mean of the three language averages; the zero-shot average includes the GlobalPIQA surprise task and the fine-tune average includes the POS surprise task. Bold marks cells where URCHIN matches or exceeds both baselines.}
\label{tab:ml}
\end{table}

\paragraph{Comparison with the official baselines.} The official per-track results are given in Tables~\ref{tab:strictsmall}, \ref{tab:strict100}, and~\ref{tab:ml}. URCHIN's advantage is concentrated in the human-like metrics: it matches or exceeds both official baselines on the Human-like Average on both strict tracks (3.71 and 3.59; Tables~\ref{tab:strictsmall} and~\ref{tab:strict100}). The gap is driven by the Reading fits, on which URCHIN's combined score exceeds both baselines on each track, together with a non-negative AoA that URCHIN decouples to zero by design whereas both baselines score negative. On the language-understanding benchmarks URCHIN trails both baselines on most tasks, as expected at 4.23M parameters, the exception being Entity-Tracking on Strict-100M, where it exceeds both (Table~\ref{tab:strict100}). On the Multilingual track (Table~\ref{tab:ml}) its 43.56 language average is above the multilingual GPT-2 baseline (35.88) and, on Chinese, above the stronger trilingual (en/nl/zh) GPT-2 baseline (43.57 against 43.31), while remaining below that baseline overall (45.94). All values of offical baselines are from the BabyLM 2026 leaderboard (11 July 2026).

\paragraph{Comparison between Parallel and Serial modes.} Parallel URCHIN and Serial URCHIN load the same weights, and the event-driven serial forward pass reproduces the parallel forward pass up to a small floating-point difference from the two integration orders (a log-domain prefix scan against a direct step recurrence). To make an item-level comparison of the difference between the two modes, both backbones are run locally on the released weights, since the official leaderboard reports only the submitted model's aggregate scores. Across the zero-shot suite (BLiMP, BLiMP-Supplement, EWoK, Entity-Tracking, COMPS, GlobalPIQA) and the fine-tuning suite, for the submitted models on all three tracks, the two implementations give the same benchmark scores: across the 66 benchmarks and tasks, together covering 445,447 individual predictions, only 90 predictions flip between the two modes, or 0.02 percent; accuracy is stable to the fourth decimal, the largest single-task difference is 0.7 points (on RTE, with 139 examples), and the Multilingual zero-shot suite is item-identical across all 170,759 of these predictions. A prediction can change only when its two candidate scores are separated by less than this floating-point difference, so it tips an argmax only on items already at the decision boundary: the two integration orders round such a near-tie in opposite directions, and only these 90 items sit that close, flipping either way and leaving the aggregate scores unchanged (Table~\ref{tab:flips}, Appendix~\ref{app:flips}). Fine-tuning is performed in the parallel mode and deployed serial for inference, so the fine-tuned scores are carried over by the same forward pass. Training and deployment are therefore the same model, not an ANN-to-SNN conversion.

\paragraph{Compute and parameter efficiency.} Training compute is counted under a convention that counts matrix-multiply operations only, treats the token embedding as a lookup at 0 FLOP, counts the LM head as a full projection, and multiplies the forward cost by three for the backward pass. Under this convention a full pretraining run costs 2.29 PFLOP on Strict-Small, 23.6 PFLOP on Strict-100M, and 18.2 PFLOP on Multilingual. The official GPT-BERT and GPT-2 baselines, counted the same way with the transformer attention term included, cost 34.6, 1009, and about 915 PFLOP on the same three tracks. URCHIN therefore reaches comparable scores using about 15, 43, and 50 times less training compute. Wall-clock training is correspondingly small, under an hour on a single H200 for every track (Table~\ref{tab:hyper}). Together with the 34K-parameter single-layer core, it is efficient in both compute and effective (non-embedding) capacity.

\begin{figure}[t]
\centering
\includegraphics[width=70mm]{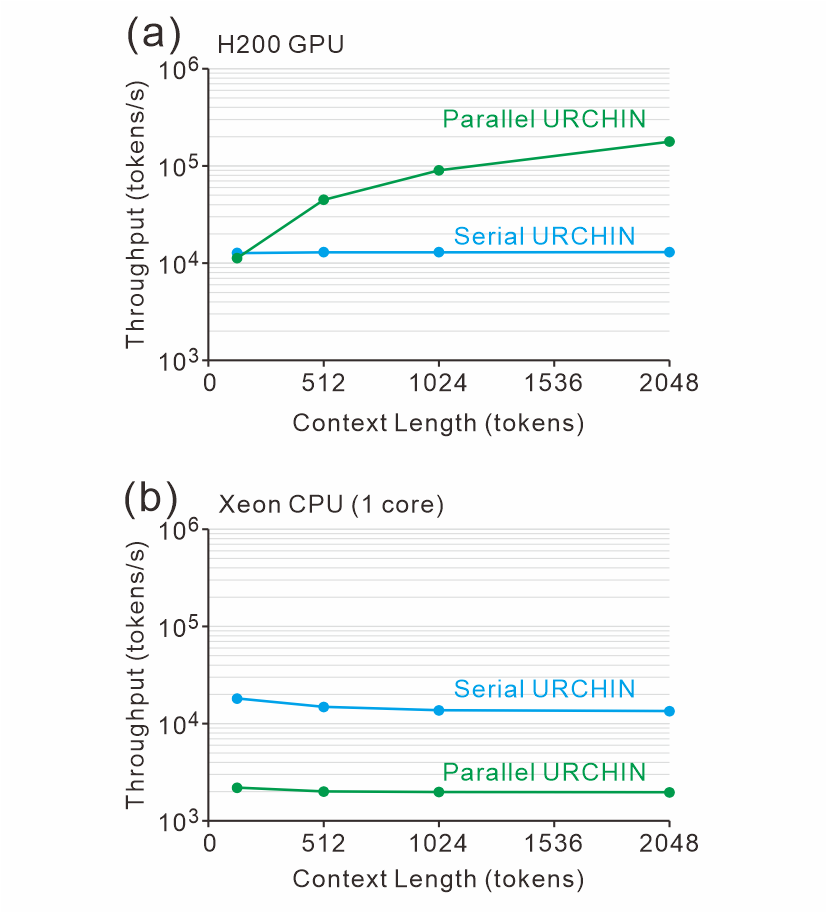}
\caption{Single-sequence inference throughput of Parallel URCHIN (24 transmission steps, evaluation setting) and Serial URCHIN (event-driven), the mean of the three homogeneous tracks. On the H200 GPU (a) parallel throughput scales with context length while serial is flat; on a single CPU core (b) the event-driven serial path is about seven times faster. Batch size one.}
\label{fig:throughput}
\end{figure}

\paragraph{Deployment throughput.} Both implementations were benchmarked on an NVIDIA H200 GPU and on a single CPU core (Intel Xeon 8480C, one thread, to emulate an edge or microcontroller compute budget), at batch size one over context lengths from 128 to 2048; the three tracks agree within about 2 percent. Because the two modes execute different amounts of work, the better mode depends on the hardware (Figure~\ref{fig:throughput}). On the GPU, Parallel URCHIN runs in near-constant time per forward pass, so its throughput grows with context length and reaches about 177,000 tokens per second at a 2048-token context, which is why it is used for training; Serial URCHIN holds a constant cost of about 13,000 tokens per second. On a single CPU core the ordering reverses: the event-driven Serial URCHIN reaches about 13,000 to 18,000 tokens per second while Parallel URCHIN falls to about 2,000, so serial is about seven times faster and is the intended edge and neuromorphic deployment path. The speedup comes from the event-driven core making a single pass instead of the 24-step transmission loop, which leaves the fixed vocabulary-projection head, rather than the spiking core, as the dominant per-token cost on a single core.

\section{Discussion}

The results bear most directly on the biological motivation for the design. The GPT-2 and GPT-BERT baselines are 12-layer transformers with 12 attention heads and 768-dimensional hidden states (official baseline model cards and leaderboard, retrieved 11 July 2026); URCHIN is a single horizontal layer of 128 spiking neurons with no attention, yet it reaches a comparable overall level and, on the human-like metrics, predicts human reading behavior better than they do (Section~\ref{sec:results}). That the smallest and hardest-constrained model is the strongest on the human-like axis is the central observation: the biological constraints act as a useful inductive bias for human-like processing rather than as a handicap, even as they leave the model behind on grammatical and world-knowledge benchmarks. The effect is clearest on the less-saturated multilingual track, a direction worth pursuing.

The two execution modes matter less at the BabyLM scale than as a deployment path that larger spiking language models could inherit. Because Parallel and Serial URCHIN are the same weights under two integration orders, such a model can be trained once at transformer-like throughput on a GPU and then run unchanged as an event-driven spiking network, with no conversion step and no drop in benchmark scores. The absence of attention keeps serial inference at a constant per-token state, with no key-value cache that grows with the sequence, so its memory does not scale with context and it runs on a single CPU core or other edge device. Its event-driven spiking form is moreover a natural match for neuromorphic hardware, which is built for sparse, per-timestep spike computation rather than dense attention. Here that path runs through a 34K-parameter recurrent core (Section~\ref{sec:results}), and the same train-parallel, deploy-serial route would carry over to larger networks.

The one human-like axis on which URCHIN does not register a signal is age of acquisition \citep{chang2022}: the official AoA correlation for the submitted model is not statistically significant. This is expected rather than surprising. URCHIN is trained with a plain causal objective and no curriculum, and in particular its data is not ordered by age of acquisition, so a strong AoA correlation is not something the training procedure is built to produce. The reading-time alignment that does appear therefore comes from the model's learned surprisals rather than from any developmental scheduling of the data; ordering the curriculum by acquisition is a separate design choice, left open here.

URCHIN is deliberately the smallest useful instance of PHCSSM: a single 128-neuron horizontal layer with no depth and no region hierarchy. This minimal choice isolates whether biological constraints can support language modeling at all under a developmental data budget, without confounding that question with capacity. Whether a wider or deeper network helps under the same data budget is left to future work. The parallel-scan formulation that makes training tractable applies equally to multi-layer and richer-cell variants of the framework. The result establishes that even a single tiny layer, under Dale's-law excitation and inhibition and lateral-only connectivity, already reaches competitive scores, which suggests these biological constraints are compatible with sample-efficient language modeling rather than an obstacle to it.

\section{Limitations}

The single small network bounds capacity, and URCHIN trails the transformer baselines on the grammatical and world-knowledge benchmarks such as BLiMP and EWoK. Learning is not part of the biological story: both pretraining and fine-tuning use standard backpropagation, so only inference runs as an event-driven spiking network, and local plasticity rules for on-device learning are left to future work. URCHIN's developmental plausibility is architectural rather than procedural: it trains on a plain causal objective, so the curriculum learning and other cognitively-motivated training strategies that much of the BabyLM community explores \citep{warstadt2023,charpentier2025} are complementary to it and remain to be combined with the spiking network. Several results come from a single training seed and are reported without a variance estimate.

\section{Conclusion}

This paper presented URCHIN, a biologically-constrained spiking language model with a Dale's-law E/I population joined only by horizontal connections, and an event-driven deployment mode, here instantiated as a single horizontal layer and applied to data-constrained language modeling across all three BabyLM 2026 tracks (Strict-Small, Strict-100M, and Multilingual). Its spiking core is a 34K-parameter single recurrent layer, pretraining costs about 15 to 50 times less than the baselines, and Parallel URCHIN and Serial URCHIN give the same benchmark scores, agreeing to the floating-point floor of the two integration orders. URCHIN provides a biologically plausible and deployable reference point in the BabyLM design space, and a base for deeper, richer-cell, and local-learning variants in later work.

\section*{Acknowledgements}
This work was supported by the National Science and Technology Council (NSTC), Taiwan (NSTC 114-2320-B-A49-027-; NSTC 114-2634-F-A49-006-; NSTC 114-2321-B-A49-003-; NSTC 114-2321-B-A49-014-). During the preparation of this work, the author used Claude (Anthropic) and Gemini (Google) for manuscript editing and code development. The author reviewed and verified all outputs and takes full responsibility for the content of this work.

\section*{Declaration of competing interest}
The author declares the following competing interests: a patent application related to the work described in this paper has been filed.

\bibliography{refs_v2}

\appendix
\onecolumn
\raggedbottom

\section{Hyperparameters}
\label{app:hyper}

\begin{table}[h!]
\centering
\small
\begin{tabular}{@{}llll@{}}
\toprule
Hyperparameter & Strict-Small & Strict-100M & Multilingual \\
\midrule
\multicolumn{4}{@{}l}{\emph{Pretraining}}\\
Corpus & \texttt{babylm\_10m} & \texttt{babylm\_100m} & \texttt{babylm\_ml\_100m} \\
Tokenizer vocabulary size & 16{,}384 & 16{,}384 & 16{,}384 \\
Context length & 2{,}048 & 2{,}048 & 2{,}048 \\
Synaptic delay & 1 & 1 & 1 \\
Sensory drive scale (\texttt{drive\_alpha}) & 0.3 & 0.3 & 0.3 \\
Transmission steps (train / eval) & 12 ($+$buffer 1) / 24 & 12 ($+$buffer 1) / 24 & 12 ($+$buffer 1) / 24 \\
Convergence-loss weight (buffer-ramp) & 800 & 800 & 800 \\
Optimizer & AdamW ($\beta_1$ 0.9, $\beta_2$ 0.95) & AdamW ($\beta_1$ 0.9, $\beta_2$ 0.95) & AdamW ($\beta_1$ 0.9, $\beta_2$ 0.95) \\
Weight decay & 0.1 & 0.1 & 0.1 \\
Peak / min learning rate & 1.5e-3 / 1e-5 & 1.5e-3 / 1e-5 & 1.5e-3 / 1e-5 \\
LR schedule & warmup $+$ cosine & warmup $+$ cosine & warmup $+$ cosine \\
Warmup steps & 500 & 2{,}600 & 2{,}000 \\
Batch size & 16 & 16 & 16 \\
Total training steps & 5{,}050 & 51{,}900 & 40{,}000 \\
Epochs & 9.78 & 10.00 & 9.93 \\
Training compute (PFLOPs) & 2.29 & 23.57 & 18.17 \\
Training wall-clock (1$\times$H200, h) & 0.08 & 0.64 & 0.48 \\
Precision & float32 & float32 & float32 \\
Random seed & 42 & 42 & 42 \\
\midrule
\multicolumn{4}{@{}l}{\emph{Fine-tuning}}\\
Tasks & GLUE (7 tasks) & GLUE (7 tasks) & 9 tasks $\times$ (en, nl, zh) \\
Optimizer & AdamW (HF defaults) & AdamW (HF defaults) & AdamW (HF defaults) \\
Learning rate & 3e-4 & 3e-4 & 1e-3 (1e-4 for POS) \\
Max epochs & 20 (WSC 30) & 20 (WSC 30) & 50 \\
Early stopping & best-per-epoch & best-per-epoch & patience 5 \\
Batch size & 32 (16 for long-seq) & 32 (16 for long-seq) & 64 (16 for POS) \\
Eval transmission steps & 24 & 24 & 24 \\
Random seed & 42 & 42 & 42 \\
\bottomrule
\end{tabular}
\caption{Pretraining and fine-tuning hyperparameters for the three tracks. The architecture (a single 128-neuron layer, 80:20 E/I, full \texttt{[1,1,1,1]} lateral connectivity, LIF neurons, no STDP) is identical across tracks. Fine-tuning tasks: the strict tracks use the seven (Super)GLUE tasks (boolq, mnli, mrpc, multirc, qqp, rte, wsc); the Multilingual track uses nine tasks (arc, belebele, bmlama, include, mnli, sib200, truthfulqa, xnli, pos), each in English, Dutch, and Chinese. The Multilingual POS fine-tuning matches the Multilingual classification fine-tuning except for the learning rate and batch size marked in POS, and additionally uses a maximum sequence length of 128 and a warmup ratio and weight decay of 0.1 and 0.01.}
\label{tab:hyper}
\end{table}

\clearpage
\section{Item-level parallel-serial agreement}
\label{app:flips}

\begin{table}[h!]
\centering
\small
\begin{tabular}{@{}llrrr@{}}
\toprule
Track & Task & N & Flips & $\Delta$ \\
\midrule
Strict-Small & BLiMP (ZS) & 59{,}875 & 4 & $+0.000033$ \\
Strict-Small & EWoK (ZS) & 7{,}618 & 1 & $-0.000131$ \\
Strict-Small & MNLI (FT) & 4{,}908 & 15 & $+0.000204$ \\
Strict-Small & QQP (FT) & 20{,}215 & 9 & $-0.000049$ \\
Strict-Small & BoolQ (FT) & 1{,}635 & 4 & $-0.001223$ \\
Strict-Small & RTE (FT) & 139 & 1 & $+0.007194$ \\
Strict-100M & BLiMP (ZS) & 59{,}875 & 4 & $0.000000$ \\
Strict-100M & EWoK (ZS) & 7{,}618 & 4 & $0.000000$ \\
Strict-100M & Entity-Tr. (ZS) & 6{,}780 & 5 & $+0.000147$ \\
Strict-100M & MNLI (FT) & 4{,}908 & 9 & $0.000000$ \\
Strict-100M & QQP (FT) & 20{,}215 & 8 & $+0.000099$ \\
Strict-100M & BoolQ (FT) & 1{,}635 & 6 & $-0.002446$ \\
Multiling. & en/SIB-200 (FT) & 201 & 1 & $-0.004975$ \\
Multiling. & en/XNLI (FT) & 2{,}000 & 4 & $0.000000$ \\
Multiling. & nl/BMLaMA (FT) & 1{,}213 & 3 & $-0.000824$ \\
Multiling. & nl/SIB-200 (FT) & 201 & 2 & $0.000000$ \\
Multiling. & zh/MNLI (FT) & 1{,}777 & 2 & $-0.000563$ \\
Multiling. & zh/XNLI (FT) & 2{,}000 & 8 & $+0.001000$ \\
\bottomrule
\end{tabular}
\caption{Item-level parallel-serial disagreements. Across the three tracks the two modes are compared on 66 benchmarks and tasks, together covering 445,447 individual predictions. \textbf{48 of the 66 are perfectly item-identical between the parallel and serial modes; only the 18 listed below disagree at all, and only on 90 of the 445,447 predictions (0.02 percent).} Each row gives the item count N, the number of flipped items, and the change in the reported score $\Delta$ (serial minus parallel); every benchmark or task not listed has zero flips and $\Delta = 0.000000$. The entire Multilingual zero-shot suite falls in the identical set: all 16 of its benchmarks, covering 170,759 of these predictions over English, Dutch, and Chinese, agree item for item. Reading is excluded, as it is scored by surprisal correlation rather than accuracy. ZS = zero-shot, FT = fine-tune.}
\label{tab:flips}
\end{table}

\end{document}